\documentstyle[sprocl,epsf]{article}

\def\Journal#1#2#3#4{{#1} {\bf #2}, #3 (#4)}

\def\NPB{{\em Nucl. Phys.} B}
\def\PLB{{\em Phys. Lett.}  B}
\def\PRL{\em Phys. Rev. Lett.}
\def\PRD{{\em Phys. Rev.} D}
\def\PRB{{\em Phys. Rev.} B}

\def\IJD{{\em Int. J. Mod. Phys.} D}
\def\IJA{{\em Int. J. Mod. Phys.} A}

\newcommand{\la}[1]{\label{#1}}
\newcommand{\be}{\begin{equation}}
\newcommand{\ee}{\end{equation}}
\newcommand{\ba}{\begin{eqnarray}}
\newcommand{\ea}{\end{eqnarray}}
\newcommand{\bi}{\begin{itemize}}
\newcommand{\ei}{\end{itemize}}
\newcommand{\rmi}[1]{{\mbox{\scriptsize #1}}}
\newcommand{\fig}{Fig.~}

\newcommand{\fr}[2]{{\frac{#1}{#2}}}

\renewcommand{\vec}[1]{{\bf #1}}

\def\lsi{\raise0.3ex\hbox{$<$\kern-0.75em\raise-1.1ex\hbox{$\sim$}}}
\def\gsi{\raise0.3ex\hbox{$>$\kern-0.75em\raise-1.1ex\hbox{$\sim$}}}
\newcommand{\lsim}{\mathop{\lsi}}
\newcommand{\gsim}{\mathop{\gsi}}

\begin{document}

\vspace*{-2cm}
\begin{flushright} hep-ph/9902282
\end{flushright}
\vspace*{1cm}

\title{MAGNETIC FIELDS AND THE EW PHASE TRANSITION%
\footnote{Presented at SEWM'98, Copenhagen, 2.-5.12.1998. 
Based on~Ref.\cite{own}.}}

\author{M. LAINE}

\address{CERN/TH, CH-1211 Geneva 23, Switzerland}

\maketitle\abstracts{ 
We review the motivation for, lattice results on, and some implications of,
external magnetic fields present at the time of the cosmological 
electroweak phase transition.}

\section{Introduction}

The existence of galactic magnetic fields today may imply the 
existence of primordial seed fields in the Early Universe. 
In order to get large enough length scales, 
it seems conceivable~\cite{revs,son} that the seed fields should have a
correlation length at least of the order of 
the horizon radius at the electroweak (EW) epoch, 
$T\sim 100$ GeV. Such large length scales could possibly be produced 
during the inflationary period of Universe expansion.\cite{inflation}

Whether or not seed fields at the horizon scale are present 
at $T\sim 100$ GeV, it is in any case clear that the existence of 
magnetic fields at somewhat smaller length scales 
cannot be excluded. Indeed, magnetohydrodynamics, 
\be
\frac{\partial \vec{B}_Y}{\partial t} = 
\frac{1}{\sigma} \nabla^2 \vec{B}_Y + 
\nabla\times(\vec{v}\times\vec{B}_Y), 
\ee
tells that magnetic fields diffuse away at scales 
$ l \lsim ({t}/{\sigma})^{1/2} \sim 
({M_\rmi{Pl}}/{T})^{1/2} T^{-1}$.
At the EW epoch this gives $l_\rmi{EW}\sim 10^7/T$. Since
diffusion continues after the electroweak epoch, 
constraints from primordial nucleosynthesis and CMBR only concern 
fields at scales larger than these, by factors
$({T_\rmi{EW}/T_\rmi{nucl.synth.}})^{1/2}\sim 10^3$ and
$({T_\rmi{EW}/T_\rmi{recomb.}})^{1/2}\sim 10^5$, 
respectively. 

A further question is the magnitude of magnetic fields. 
An equipartition argument would say that only a small 
fraction of the total (free) energy density can be in magnetic 
fields. This leads to $B_Y/T^2 \lsim 2...3$.  

In conclusion, there could well be essentially homogeneous and 
macroscopic ($l_\rmi{EW}\gg T^{-1}$) 
magnetic fields around at $T\sim 100$ GeV, with 
a magnitude $B_Y/T^2 \sim 1$. 
The purpose of this talk is to review how 
such fields would affect the cosmological EW  phase transition. 


\section{The perturbative EW phase diagram in an external field}

To be specific, we study here a system with a fixed
hypercharge flux $\Phi_{B_Y} = \int d\vec{s} \cdot \vec{B}_Y$.
The properties of a system with a fixed field strength can be 
obtained with a Legendre transformation. 
The thermodynamics 
depends essentially on three dimensionless
parameters, 
\be
x 
\sim \fr18 \frac{m_H^2}{m_W^2}, 
\quad 
y 
\sim 4.5 \frac{T-T_0}{T_0}, 
 \quad
b \sim 2.0 \frac{\Phi_{B_Y}}{(\mbox{area})}\frac{1}{T^2},
\ee
where $T_0$
equals the critical temperature up to radiative corrections and 
the numerical factors are there for technical reasons. 
Thus, one can tune $x$ (Higgs mass), $y$ (temperature) and $b$ 
(magnetic flux density). A further parameter, $z \sim \tan^2\theta_W$,
is fixed to a constant value $z\approx 0.3$.

\begin{figure}
 
\epsfxsize=6cm
\centerline{\epsffile{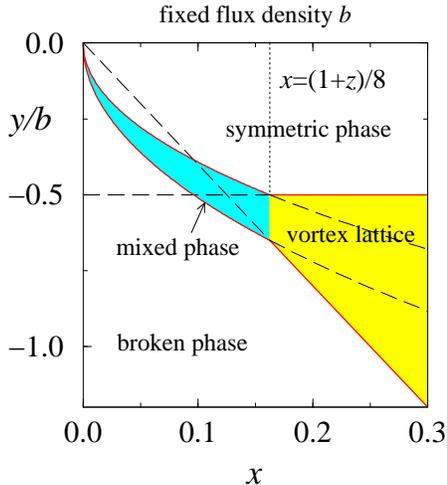}}
 
\vspace*{-0.7cm}

\caption[a]{The tree-level phase diagram of the electroweak
theory in a magnetic flux.\la{tree_b}}
\end{figure}

In the space of these variables, the tree-level phase diagram of
the theory is as shown in \fig\ref{tree_b}. At small $x$, there is 
a first order phase transition,\cite{gs} the stronger the larger $b$.
(1-loop corrections~\cite{eek} 
do not change this statement in any essential way.) 
In the present ensemble the transition takes place through a mixed phase,
containing simultaneously macroscopic regions of the symmetric
and broken phases. 
At large $x$, the situation is different: the ground state solution 
of the classical equations of motion is a vortex lattice, 
or the ``Ambj{\o}rn-Olesen phase''.\cite{ao} 

\section{The non-perturbative phase diagram}

The perturbative phase diagram discussed 
above need not be reliable at large Higgs masses, $x\gsim 0.1$, 
since the dynamics of the theory is completely non-perturbative
there.\cite{endpoint} Thus, 
the existence of a vortex lattice 
is also not obvious. 

It is perhaps illuminating to consider more generally whether 
a vortex lattice phase can really exist in three dimensions
in a fluctuating system. We will skip the question of whether 
such a phase can exist in a strict mathematical sense, and 
consider the issue from a more practical point of view.
Then, it is clear that a lattice phase could exist in principle,
since an Abrikosov vortex lattice and a related 1st order melting
transition have been observed even experimentally 
in a very similar system, a superconductor. 
There is a phenomenological test,
called the Lindemann criterion, for when a lattice
phase can be observed: the requirement is that the root mean 
square fluctuation of the flux line position around its average, 
be sufficiently smaller than the lattice spacing.\cite{lindemann} 
In superconductors this criterion can be satisfied, since the 
fluctuations can be reduced by going to lower and lower temperatures. 
In the present system the requirement is more difficult to 
satisfy, since, for fixed $b$, one cannot go to arbitrarily 
low temperatures within the vortex phase, see Fig.~1. However, 
a quantitative estimate of the average fluctuations is lacking 
at the moment, and we thus have to turn to lattice experiments. 


A fixed magnetic flux can easily be implemented on 
the lattice, by using a 3d effective field
theory~\cite{generic,su2u1,mssm}
and by modifying the boundary conditions
related to the hypercharge U$_\rmi{Y}$(1) field:
\be
\Phi_{B_Y} = \int d\vec{s} \cdot \vec{B}_Y = 
\int dx_1dx_2 F_{12} = \oint d s_i A_i. 
\ee
Otherwise the simulations proceed very much like without
a magnetic field.\cite{su2u1}

There is one important new lattice artifact to be mentioned. 
Indeed, with the ensemble used, there can be a mixed
phase in the system, as shown in Fig.~1. A mixed phase 
implies the existence of surfaces, which cost energy. Only 
at large enough volumes does the bulk free energy win over the 
surface energy.
Thus, for finite lattice sizes, small 
values of $x$, corresponding to a strong 1st order transition
and large surface tension, do 
not display a mixed phase.\footnote{We thank M. Tsypin for 
discussions important for our appreciation of this issue.} 
For large
$x$, $x\ge (1+z)/8$, on the other hand, the surface energy 
becomes negative as in type II superconductors and the 
strong volume dependence disappears. Hence, these volume 
artifacts do not appear in the vortex lattice phase.  


\begin{figure}
 
\epsfxsize=6cm
\centerline{\epsffile{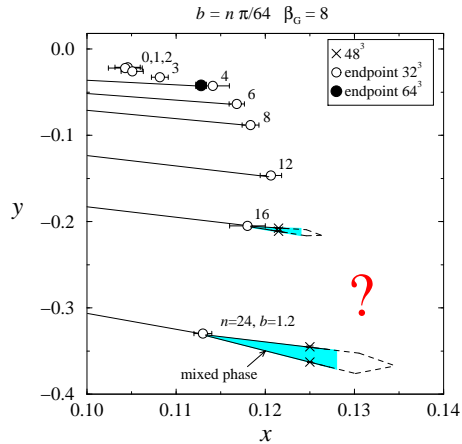}}
 
\caption[a]{The non-perturbative phase diagram of the electroweak
theory in a magnetic flux, for finite volumes. 
The behaviour in the region of the question mark is still
unknown.}
\end{figure}

The basic results of the lattice simulations are as follows:

{\bf First order regime}, $x<(1+z)/8$: the transition is 
qualitatively similar to that in perturbation theory, Fig.~1:
a strong 1st order transition. However, for a given lattice volume,
one has to go to a large Higgs mass where the surface 
tension is sufficiently small to see the mixed phase. 
A phase diagram determined from finite volume 
lattices is shown in Fig.~2. 

%
%

{\bf Intermediate regime}, $x\sim (1+z)/8$:  when $x$ 
is increased, the phase transition gets weaker 
and finally ends, and the endpoint
looks qualitatively similar to the one 
described in Ref.\cite{endpoint},
in the absence of a magnetic field.

{\bf Vortex lattice regime}, $x>(1+z)/8$: for the magnetic fields
studied so far, we have not observed the vortex lattice phase! 
The individual or averaged 
configurations do not display any non-trivial
structure, and there are no qualitative changes in any of the 
observables measured. 


\nopagebreak

\section{Conclusions}

We have seen that 
an external magnetic field makes 
a 1st order transition 
stronger, but there is still no transition in the Standard Model
for $x\gsim 0.14$ (at least for $B_Y/T^2 \lsim 0.5$), 
corresponding to $m_H\gsim 90$ GeV. Larger 
fields have not yet been conclusively studied. 
We have not observed a vortex lattice phase 
with qualitatively new properties at these magnetic 
fields, even though the solution of the classical equations
of motion displays such a phase. 

In the MSSM, there can be a 1st order transition for the 
experimentally allowed Higgs masses. Then an external magnetic 
field might have implications, e.g., for baryogenesis:
the bubble dynamics and the real time history of the transition
get modified, and the external magnetic field can couple also 
directly to fermion number non-conservation.\cite{gs}

\section*{Acknowledgments}

The work reported here was done in collaboration with K. Kajantie, 
J. Peisa, K. Rummukainen and M. Shaposhnikov (Ref.\cite{own}). 
I am also grateful to  P. Pennanen, 
A. Rajantie and M. Tsypin for very useful discussions. This work 
was partly supported by the TMR network {\em Finite Temperature Phase
Transitions in Particle Physics}, EU contract no.\ FMRX-CT97-0122.


\end{document}